\documentclass[a4paper,pdflatex]{article}
\usepackage{graphicx,amsmath,amsfonts}
\begin{document}
\vspace*{.25cm}

\large

\noindent {\Large \bf Coherent production of the long-lived pionium $nP$ states
in relativistic nucleus--nucleus collisions}

\vspace{.5cm}

\hspace{2.5 cm}
S Gevorkyan\footnote{
gevs@jinr.ru}
and O Voskresenskaya\footnote{voskr@jinr.ru}

\vspace{.5cm}

\parbox{13cm}{
{\small Joint
Institute for Nuclear Research, Dubna, Moscow Region, 141980 Russia}}

\begin{abstract}

The coherent production of the $nP$ states of the $\pi^+\pi^-$  atoms ($A_{2\pi}$) in
relativistic nucleus--nucleus collisions is considered as a possible source of the
$A_{2\pi}(nP)$ beam for the pionium Lamb-shift measurement. A general expression for
estimation  of the $A_{2\pi}(nP)$ yield is derived in the framework of the equivalent
photon approximation.

\end{abstract}

\rm
\section{Introduction}
The  DIRAC  experiment  \cite{Dirac14,dirac} aims to  observe  and  study
hydrogen-like atoms formed by pairs of  $\pi^+\pi^-$ and  $\pi^\pm K^\mp$ mesons
($A_{2\pi}$  and $A_{\pi K}$ atoms, respectively) that produced in their ground-states
in inclusive processes with the use of the 24 GeV proton beam at PS CERN.
The ground-state lifetime measurement of these atoms allows to obtain in a model
independent way the difference of $\pi\pi$ ($a_0-a_2$) and
$\pi K$ ($a_{1/2}-a_{3/2}$) scattering lengths in $S$ state. The final measurements of the
ground-states $A_{2\pi}$ and $A_{\pi K}$ lifetimes gave the values
$\tau=3.15\cdot 10^{-15}$ s and $\tau=2.5\cdot 10^{-15}$ s, respectively \cite{Dirac14}.

The  lifetimes  of  the dimesoatom  $nP$  states  are
3--5  orders of magnitude higher in comparison  with  the ground-state lifetimes,
since for for these states the strong interaction is suppressed. As a result, both the lifetime and mean path  of such  states are  a  few  order  higher  compared to the ground state.
Observation of such long-lived (meta-stable)
states of $A_{2\pi}$ opens the possibility to measure the Lamb-shift in $A_{2\pi}$
and also to obtain the other combination the $\pi\pi$ scattering lengths: $2a_0+a_2$.
DIRAC plans to study the Lamb shift  in  $A_{2\pi}$   atoms
and then to extract the latter combination of the scattering lengths.

The measurement the energy  splitting between levels $nS$ and $nP$ coupled with the
lifetime measurement would provide a determination of $a_0$ and $a_2$  separately.
The values of these scattering lengths can be rigorously calculated in Chiral Perturbation
Theory (ChPT) \cite{3,4}. Thus, these measurements
provide an experimental test of the low-energy QCD  predictions.

Some possibilities for measuring the pionium Lamb shift $\Delta E_n=E_{nS}-E_{nP}$
based  on observation  of  the interference between $nS$ and $nP$ ($m = 0$) states in the
external electro-magnetic fields are discussed in the papers \cite{nemen}.
For this aim the beams of relativistic $A_{2\pi}(nP)$ are needed. One way of obtaining of
such beam is regarded in \cite{2015}. An another way was proposed in \cite{Bern}.

In the present work, we consider the coherent production of the long-lived pionium $nP$
states in relativistic nucleus--nucleus collisions as the possible source of $A_{2\pi}(nP)$
beam for the pionium Lamb-shift measurements.

\section{Primary idea}
Let us consider the production of $nP$ states of pionium atoms in the coherent
collisions of the projectile $A_P$   and target $A_T$ nuclei
\begin{eqnarray}
 A_P + A_T  \rightarrow A_P   + A_T  + A_{2\pi}(nP).
\end{eqnarray}

    Since the quantum numbers of pionium in $nP$ states are the same as photon's ones, the
coherent photoproduction process
\begin{eqnarray}
\label{2}
 \gamma + A_T   \rightarrow A_{2\pi}(nP)+ A_T
\end{eqnarray}
is quite intensive in GeV region.  Another advantage of the coherent reaction (\ref{2})
is the sharp angular distribution  the produced pionium, comparable with angular
acceptance of experimental setup.

The  source  of  "effective"  photons  could  be  provided  by  projectile  nucleus
$A_P$ in the spirit of EPA (Equivalent Photon Approximation) \cite{bert}.

\section{Reaction amplitude}

The amplitude of (\ref{2})  can be calculated as a projection of the amplitude of the
reaction
\begin{eqnarray}
\label{3}
 \gamma + A_T   \rightarrow \pi^+  + \pi^- + A_T
\end{eqnarray}
on the $nP$ state of pionium.

The last one may be represented by following Feynman diagrams:

\begin{figure}[h!]

\begin{center}

\includegraphics[width=0.65\linewidth]{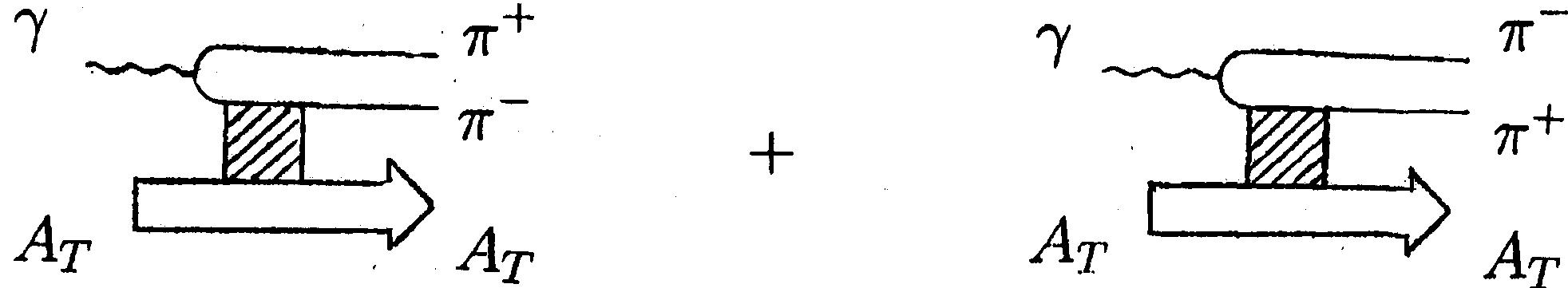}

\caption{The  Feynman diagrams relevant the reaction (\ref{3}).
} \label{Fig1}

\end{center}

\end{figure}

\newpage
\noindent Here, the boxes represent the amplitudes of the $\pi^\pm A_T$ scattering.

The latter amplitude reads
$$M\Biggl(\gamma + A_T   \rightarrow \pi^+  + \pi^- + A_T\Biggl)$$
\begin{equation}
=e\vec\epsilon\vec p\left[\frac{M_{\pi^+A_T}(\vec q)}{D_1}+
\frac{M_{\pi^-A_T}(\vec q)}{D_2}\right],
\end{equation}
\begin{equation}
D_1=m^2_\pi+ \vec k^2_{1T},\qquad D_2=m^2_\pi+ \vec k^2_{2T},
\end{equation}
\begin{equation}
\vec p= \frac{\vec k_{1T}- \vec k_{2T}}{2},\qquad
\vec q=- \frac{\vec k_{1T}+ \vec k_{2T}}{2},
\end{equation}
where $e$ is the elementary charge, $\vec\epsilon$ is the photon polarization vector,
$\vec k_{1T}$ and $\vec k_{2T}$ are the transverse momenta of $\pi^-$ and $\pi^+$,
respectively.

For targets with zero value of isotopic spin ($T=0$)
$M_{\pi^-A_T}= M_{\pi^+A_T}$.
In general case, $M_{\pi^-A_T}$ and  $M_{\pi^+A_T}$ differ only slightly. So, we put
\begin{equation}
M_{\pi^-A_T}= M_{\pi^+A_T}\equiv M_{\pi A_T}.
\end{equation}

In this approximation, the final result reads
$$
 \frac{d\sigma}{d\omega}\Bigg(A_P + A_T  \rightarrow A_P   + A_T  + A_{2\pi}(nP)\Bigg)
$$
\begin{equation}
\label{8}
=\frac{n_{eff}\left(\omega , Z_p\right)}{\omega}\cdot \alpha^6\cdot \sigma^{el}_{\pi A_T}
\left(\frac{1}{n^3}-\frac{1}{n^5}\right),
\end{equation}
where  $n_{eff}$  is  the  number  of  "effective"  photons \cite{bert} with  the  energy
$\omega  =  E_{\gamma}=E_{A_{2\pi}}$ produced  by projectile nucleus with charge $Z_P$,
and $n$ is the principal quantum number of the pionium $nP$ state. This expression can
be used for the estimation of the $A_{2\pi}(nP)$ yield.

We consider this paper  as a preliminary report on the obtained results.

\section*{Acknowledgments}

The authors are grateful to Leonid Afanasyev for useful comments.
This work is devoted to the memory of our friend and collaborator
Alexander Tarasov.

\end{document}